**Mathematical modelling indicates that lower activity of the haemostatic system in neonates is primarily due to lower prothrombin concentration**

**Ivo Siekmann**[1,2,*], **Stefan Bjelosevic**[3,4], **Kerry Landman**[5], **Paul Monagle**[3,6,7], **Vera Ignjatovic**[3,7,†] **& Edmund J. Crampin**[1,2,5,8,†]

[1]Department of Applied Mathematics, Liverpool John Moores University, England
[2]Systems Biology Laboratory, University of Melbourne, Australia
[3]Haematology Research, Murdoch Children's Research Institute, Melbourne, Australia
[4]The Sir Peter MacCallum Department of Oncology, University of Melbourne, Australia
[5]School of Mathematics and Statistics, University of Melbourne, Australia
[6]Department of Clinical Haematology, Royal Children's Hospital, Melbourne, Australia
[7]Department of Paediatrics, University of Melbourne, Australia
[8]School of Medicine, University of Melbourne, Australia

[†]Vera Ignjatovic and Edmund J. Crampin contributed equally to this study.

Corresponding Author:
[*]Ivo Siekmann, Department of Applied Mathematics, Faculty of Engineering and Technology, Liverpool John Moores University, James Parson Building, 3 Byrom Street, Liverpool L3 3AF, England, phone: +44 151 231 2092, e-mail: i.siekmann@ljmu.ac.uk

**Haemostasis is governed by a highly complex system of interacting proteins. Due to the central role of thrombin, thrombin generation and specifically the thrombin generation curve (TGC) is commonly used as an indicator of haemostatic activity. Functional characteristics of the haemostatic system in neonates and children are significantly different compared with adults; at the same time plasma levels of haemostatic proteins vary considerably with age. However, relating one to the other has been difficult, both due to significant inter-individual differences for individuals of similar age and the complexity of the biochemical reactions underlying haemostasis. Mathematical modelling has been very successful at representing the biochemistry of blood clotting. In this study we address the challenge of large inter-individual variability by parameterising the Hockin-Mann model with data from individual patients, across different age groups from neonates to adults. Calculating TGCs for each patient of a specific age group provides us with insight into the variability of haemostatic activity across that age group. From our model we observe that two commonly used metrics for haemostatic activity are significantly lower in neonates than in older patients. Because both metrics are strongly determined by prothrombin and prothrombin levels are considerably lower in neonates we conclude that decreased haemostatic activity in neonates is due to lower prothrombin availability.**

INTRODUCTION

Recent experimental studies have shown that the plasma levels of blood clotting proteins vary considerably, both with age, as well as between healthy individuals of the same age[1]. The haemostatic system in neonates and children also shows considerably different functional characteristics compared to adults[2-4]. Therefore, clinical studies of the adult haemostatic system may not be transferrable to the treatment of thrombotic and haemorrhagic disorders in young patients.

The high variability of blood clotting protein abundances is an intrinsic property of the haemostatic system rather than being caused by flaws of the available experimental methods. The complexity of the haemostatic system in combination with the strong fluctuations of protein concentrations between individuals and with age makes determining the impact of experimentally measured differences in plasma levels of individual haemostatic proteins extremely challenging. Mathematical modelling is a useful tool for investigating the relative importance of a variety of complicated interdependencies in complex systems, and has also been successfully applied to the haemostatic system. Early qualitative models focused on investigating the mechanism of haemostatic response in rapidly forming a blood clot shortly following an injury[5,6]. More recently, this behaviour known as excitability was investigated in mathematical models by Jesty, et al.[7] and Beltrami and Jesty[8].

The next generation of haemostasis models was based on more detailed representations of the biochemical reaction network of haemostatic proteins. Representing chemical reactions by mass-action kinetics allows for a more data-driven approach by measuring rate constants experimentally. Thus, given that sufficient experimental data is available, these models not only capture qualitative aspects such as excitability but may also be used to study haemostasis quantitatively. The model by Hockin, et al.[9], henceforth referred to as the Hockin-Mann model, consists of a system of 34 ordinary differential equations that represents the dynamics of 44 biochemical reactions of the haemostatic network by mass action kinetics. Each equation represents the time-dependent dynamics of one of the haemostatic proteins, or a complex of proteins. For a given set of initial concentrations, the Hockin-Mann model enables us to calculate how these haemostatic factors change over time.

The Hockin-Mann model continues to be used as the starting point or a building block in new models that are being developed. Wajima et al. extend a previous version of the Hockin-Mann model in order to study several experimental tests of the haemostatic system, its response to different treatments (warfarin, heparin, vitamin K) and to perturbations by Taipan snake venom[10]. Models based on the Hockin-Mann model have been developed for studying the effect of anti-coagulants such as e.g. the novel anti-coagulant (NOAC) rivaroxaban[11,12]. Most recently Mitrophanov et al. used the Hockin-Mann model for investigating the effects of acidosis on thrombin generation[13].

Gaining systematic insight into the behaviour of highly detailed models within the order of tens or even hundreds of equations is a difficult problem. Danforth et al. follow a computational approach for investigating the parameter sensitivity of the Hockin-Mann model[14]. In this study they examine the sensitivity of thrombin generation as well as the behaviour of the full model on variations of individual rate constants in a range from 10 to 1000 % of the reference values given in the original publication[9]. In Danforth, et al.[15] the statistical approach from[14] is used to assess the sensitivity of thrombin generation to variations in blood haemostatic factors, both utilising synthetically generated combinations of concentrations, as well as data from healthy adults (patients with haemophilia A and a population receiving warfarin). The study presented here follows a similar approach as Danforth, et al.[15] but we investigate healthy individuals from different age groups rather than different adult populations. Most recently, Dunster and King[16] demonstrated, via a thorough mathematical analysis of one of the earlier models of haemostasis by Willems, et al.[17], how, taking advantage of distinct time scales, simplified models for different phases of thrombin generation can be derived. These phases can be related to the initiation, propagation and termination phase of thrombin generation. Because the simplified models are much easier to analyse than the full model, the dominant mechanisms in each of six different phases can be clearly identified.

This study aims to investigate age-dependent changes of the blood clotting system. In the absence of knowledge regarding changes to the network of biochemical reactions itself, we use the Hockin-Mann model as a representation of the haemostatic system of all age classes. In order to test the hypothesis that age-dependent changes are due to differences in the plasma levels of blood clotting proteins we use age-stratified data by Attard, et al.[1]. In contrast to previous studies we explicitly account for variability between individuals by parametrising the initial concentrations of haemostatic factors with data from individual patients – in fact, our results show that choosing the levels of haemostatic factors based on data aggregated for different age groups would be misleading.

Different to earlier experimental studies of the age-dependent blood clotting system which were based on functional assays[2,3], Attard et al. have presented the first comprehensive data set that provides quantitative age-stratified protein levels. Quantitative assays are clearly more suitable than functional data for obtaining blood clotting factor concentrations necessary for parametrising models of haemostasis because functional assays only provide a proxy for concentrations. For this reason, caution is required when comparing with older data sets based on functional assays.

To the best of our knowledge, this is the first time that a model of the haemostatic system is parameterised with age-stratified haemostatic protein abundance data. This allows us to link observed age-dependent differences in the concentrations of individual haemostatic factors to functional implications for the activity of the haemostatic system.

RESULTS

In order to account for the strong inter-individual variability in different age groups we parametrised the initial protein concentrations in the Hockin-Mann model with levels of prothrombin, FV, FVII, FVIII, FIX and FX measured in individual patients[1]. For all age groups, we compared TGCs calculated from the mean levels of these haemostatic proteins, see Supplementary Material, Figure S1, with TGCs calculated for individual patients (Figure S2). In analogy to experimental studies, four indices were used for quantifying different aspects of a given thrombin curve (Figure 2): The lag time (LAG) – the time it takes until thrombin exceeds a certain fraction of its maximum - and the time it takes until thrombin reaches its maximum, the time to thrombin peak (TTP) indicate how quickly thrombin is generated in response to stimulation by TF. The other two indices, the maximum total thrombin generation (MAX) and the area under the thrombin curve (AUC), are measures for the strength of this response.

The results show that a TGC calculated for the mean plasma levels of haemostatic proteins of a particular age class fails to accurately represent some of the properties of the TGCs calculated for individual patients. For example, whereas the TGCs calculated from mean plasma levels suggest that the TTP is approximately $t = 300s$, especially the results for teenagers and adults suggest that no such pattern exists because the TTP strongly varies within age groups. It was therefore considered essential to carry out further analysis based on TGCs calculated for individual patients. The results for individuals in each of the age groups for LAG, TTP, MAX and AUC are presented in Figure 3. The levels of LAG and TTP show no significant differences between different age groups. In contrast, both thrombin maximum (MAX) and area under thrombin curve (AUC) exhibit a significant increase from a low level observed for neonates at day 1 and day 3 compared to other age groups. Also, the higher value of MAX and AUC for children less than one year old until adults is statistically similar (Figure 3).

We then investigated the relationship between MAX and AUC and the initial concentrations of the blood clotting factors reported by Attard, et al. [1]. Figure 4 shows that MAX appeared to increase linearly with prothrombin concentration. Figure 5 suggests that AUC was solely determined by the concentration of prothrombin and was not dependent of the concentrations of all other haemostatic factors.

Age-related differences in the concentrations of haemostatic factors, based on Attard, et al. [1] are shown in Figure 1. The concentrations of factor VII and factor IX clearly increased with age. The concentration of factor VIII did not change with age whereas for concentrations of factors V and X the dependence with age is less obvious. However, for the concentration of prothrombin we observed a clear jump from a low level of approximately 0.5 IU/mL to more than 1.0 IU/mL in the other age groups. Hence there were essentially two different levels of prothrombin concentrations for neonates compared with the other age groups, in contrast to factors VII and IX whose concentrations increased continuously with age.

DISCUSSION
Developmental haemostasis is a concept that describes multiple and complex age-specific differences in the haemostatic system. The overall impact of these differences is believed to provide protection for the neonate and child in terms of response to bleeding and clotting stimuli. However, predicting the response to diseases that affect haemostasis, or to the multitude of new anticoagulant drugs that are becoming available for clinical care in children is challenging. Conducting large-scale trials of these drugs in children, such as those performed in adults, has proven to be very difficult. Thus, the development of a mathematical model that simulates the age appropriate haemostatic system would be very advantageous.

We presented a novel mathematical modelling approach that enables us to link age-related differences of haemostatic factors in neonates, children and adults to functional differences in their haemostatic system. The key contribution of this study is that data from individual patients has been used for parametrising the Hockin-Mann model. This enables us to appropriately account for the strong inter-individual variability present in different age groups, in contrast to the more common approach of obtaining aggregated parameters from population statistics – compare Figures S1 and S2 in the Supplementary Material. Because we calculate a thrombin curve for each individual patient, unlike previous studies, we are in a unique and novel position to draw conclusions regarding the variability of haemostatic activity based on realistic distributions of blood clotting factors from individual patients. We note that this is different to simply investigating the effect of upper and lower bounds of individual blood clotting factors because this ignores correlations between different factor levels – the strong variations of the correlation structure between various haemostatic factors is shown in Figure S3 of the Supplementary Material.

Our approach is based on the hypothesis that the biochemical reactions of the haemostatic system are similar between different subjects and remain unchanged with age so that differences in haemostatic factor concentrations are the main source of observed age-related differences in haemostatic activity. Under these considerations, a model parameterised with age-stratified abundances of haemostatic factors (while leaving the reaction rate constants unchanged) fully accounts for age-related differences in the haemostatic system. The implications of these assumptions, in particular, the role of inhibitors, will be discussed in more detail below. Moreover, we will present a preliminary data set collected by two of the co-authors that confirms the qualitative differences between neonates and older age groups predicted by the model.

Our main finding is that for neonates at day 1 and day 3 post-birth, MAX and AUC were significantly lower than in older age groups indicating a lower activity of the haemostatic system. Importantly, our model enables us to identify a significantly lower prothrombin level as the most likely cause for this observation. This simple relationship between prothrombin and indicators of blood clotting activity bypasses the complex correlation structure between haemostatic factors apparent in Figure S3 of the Supplementary Material.

First, we found that whereas LAG and TTP are unchanged with age, MAX and AUC are significantly different between neonates and all other age groups. Both are elevated from a low level of approximately 200 nM or 500 nM·min, respectively, for neonates at day 1 and day 3 to significantly higher levels of 400 nM or 1000 nM·min for all other age groups. In Figure 6 we demonstrate that the increase of AUC for age groups older than day 3 predicted by the model is consistent with age-stratified measurements of AUC (Ignjatovic and Monagle, unpublished). This confirms that although some possible age-dependent changes in the haemostatic system are not accounted for, the model behaviour related to the indices considered in this study is nevertheless qualitatively correct.

Second, we observed that prothrombin concentration is a strong predictor for MAX and AUC. This has been reported in both the modelling[15] as well as in the experimental literature[18] albeit in different settings as our study. Danforth, et al. [15] showed that both MAX and AUC depend most sensitively on prothrombin in their simulation study of the Hockin-Mann model. For the Willems, et al. [17] model, Dunster and King [16] demonstrate that both MAX as well as AUC increase linearly with prothrombin concentration and give Duchemin, et al. [18] as a reference which experimentally confirms this observation from the mathematical model. None of the above studies refers to different age groups.

Third, by combining our observations, we identify the lower prothrombin concentration as the cause for the lower values of MAX and AUC in neonates. Whereas our model shows that MAX and AUC are lower in neonates at day 1 and day 3 post-birth than in the older age groups, we know from the Attard, et al. [1] data that neonates also have a decreased prothrombin concentration compared to adults. Taking into account that MAX and AUC are largely determined by prothrombin we conclude that the lower thrombin production in neonates is primarily due to lower prothrombin availability. Although a strong influence of prothrombin and indices related to thrombin abundance is not unexpected it is interesting that the model suggests that other blood clotting factors hardly play any role in determining MAX and AUC. Given that the haemostatic network is in general characterised by a high level of complexity this observation is potentially very useful because it suggests that targeting prothrombin may to a great extent be sufficient for regulating the amount of thrombin generated.

Because our study relies strongly both on data from individuals as well as quantitative rather than functional assays, the influence of several blood clotting proteins such as the inhibitors Tissue Factor Pathway Inhibitor (TFPI) and Antithrombin as well as Fibrinogen and thrombomodulin could not be accounted for because they were not measured by Attard, et al. [1]. To the best of our knowledge, comparable age-stratified data sets for these blood clotting factors based on quantitative assays are currently not available. Moreover, although Attard et al. measured Factor XI (FXI), FXII as well as protein C and protein S, these blood clotting proteins could not be included because they are not accounted for by the Hockin, et al. [9] model. Before we discuss the possible roles of different inhibitors, we emphasise that our approach to age-dependent modelling can be easily applied to an extended model as more comprehensive data becomes available.

Andrew, et al. [19] observe that the inhibitor Antithrombin was decreased in neonates by 60% and it has been suggested that this enabled normal haemostatic function. But the role of inhibitors in the neonate haemostatic system remains controversial because whereas Antithrombin is decreased, the inhibitor $\alpha_2$ Macroglobulin ($\alpha_2$M) is increased. In fact, our results are consistent with a recent article by Kremers, et al. [20] which attributed the decreased thrombin generation in young patients to decreased prothrombin conversion and to a lesser extent to elevated levels of $\alpha_2$M. Nevertheless, the issue of clarifying the relative importance of the different inhibitors in neonates clearly deserves further investigation. Because this requires extending the Hockin-Mann model by the inhibitor $\alpha_2$M we leave the study of this interesting question to future work.

In this study, we aimed to represent the experimental assay for determining the activity of the haemostatic system via measurement of the thrombin generation curve. For representing the *in vivo* system more detailed models of haemostasis have been developed that account for the fluid dynamics of blood within the constraints of the blood geometry as well as the interaction with platelets, see, e.g. Cito, et al. [21] or the relevant chapters in Ambrosi, et al. [22].

In summary, we have demonstrated that parametrising a mathematical model with data of individual patients from different age groups is a useful tool for investigating age-dependent differences in the haemostatic system. As additional data become available, our approach can be easily transferred to more comprehensive models of the *in vitro* or *in vivo* haemostatic system which will enable us to explore the complex system underlying developmental haemostasis in more detail. Considering the considerable uncertainty regarding treatment of young patients with coagulation disorders, an extremely useful direction will be to incorporate the interactions with haemostatic drugs in the age-stratified model presented here.

METHODS

In order to account for the strong variability between individuals apparent in the data by Attard et al., we simulated the Hockin-Mann model with different parameters for each patient sample. This approach has been used earlier for investigating inter-individual variability in adults[15]. Based on these simulation results we then quantified the activity of the haemostatic system by four commonly used indices that are derived from the simulated thrombin concentration over time, the so-called thrombin generation curve (TGC): the lag time (LAG), the time to thrombin peak (TTP), the maximum thrombin concentration (MAX) and the area under the thrombin curve (AUC). Finally, we investigated which individual haemostatic factors had the strongest influence on each of the four metrics and related our results to differences in the availability of haemostatic factors between different age groups.

*Parameterisation:*
We parameterised the Hockin-Mann model using age-stratified data from the study by Attard, et al.[1], that consists of measurements for factors II and V given as international units (IU/mL) and factors VII, VIII, IX and X reported as percentages (%). Data for these six haemostatic factors were available for patients from seven age groups, 10 samples each for neonates at day 1 and day 3 of age and 20 samples each for the remaining age groups (younger than 1 year, 1-5 years, 6-10 years, 11-16 years, adults). The distributions over the individual age groups are summarised in Figure but note that in this study we used individual samples – i.e. the raw data. Measurements were converted from the original units to concentrations by choosing the average adult results measured by Attard et al. as a reference parameter set, see Table 1. We required that these average adult results correspond to the initial concentrations of the Hockin-Mann model, see Table III in Hockin, et al.[9] or Table 1, which were chosen as average values in human plasma. Thus, measurements from an individual subject were converted to concentrations by scaling with respect to the average result for adults, as reported in Attard, et al.[1].

*Simulation:*
For each parameter set, our own implementation of the Hockin-Mann model was simulated using a numerical solver of the GNU Scientific Library (GSL)[23]. For $t = 0$ initial conditions for factors II, V, VII, VIII, IX and X were chosen according to a sample from the data by Attard, et al.[1] as described previously. Thrombin generation was initiated by initialising Tissue Factor (TF) at [TF]=5pM. The model was simulated until $t_{end} = 5000s$ to ensure that thrombin generation proceeded through all phases from initiation via propagation to termination. Total thrombin is represented by two fractions in the Hockin-Mann model, meizothrombin (mIIa) and thrombin (IIa). Due to the higher activity of meizothrombin, the total amount of thrombin over time, the thrombin generation curve (TGC), is calculated by [IIa]*(t)* + 1.2 [mIIa]*(t)* – for details, see Hockin, et al.[9]. A representative example (where termination occurred approximately at $t = 1400$ s) is shown in Figure 2.

*Indices of thrombin generation:*
In analogy to experimental studies, thrombin generation was characterized quantitatively by four metrics: The lag time (LAG) and the time to thrombin peak (TTP) indicate how quickly thrombin is generated in response to stimulation by TF. In contrast, the remaining two indices, the maximum total thrombin generation (MAX) and the area under the thrombin curve (AUC), are measures for the strength of this response. The maximum total thrombin concentration is defined as

$$\text{MAX} := max_{t \in [0, t_{end}]}\big([\text{IIa}](t) + 1.2[\text{mIIa}](t)\big)$$

where $[\text{IIA}](t)$ and $[\text{mIIa}](t)$ are thrombin and meizothrombin concentration over time, respectively, and $t$end is the end time of the simulation. The time to thrombin peak (TTP) is then defined as the elapsed time after stimulating the system with tissue factor until the time the concentration MAX is attained. Similarly, the lag time LAG is defined as the time until 1/6 of the maximum thrombin concentration MAX is reached. The area under the thrombin curve $AUC$ is given by the integral over the thrombin curve

$$\text{AUC} := \int_0^{t_{end}} [\text{IIa}](t) + 1.2[\text{mIIa}](t)\, dt$$

All four indices are visualised in Figure 2.

*Analysis of age-stratified simulation results:*
All statistical analyses of the simulation results were performed in the R software[24]. From the simulations based on the age-stratified data from Attard et al. we obtained age-stratified results for the four indices LAG, TTP, MAX and AUC. Differences between age groups were determined by comparing the distributions of each index. For indices that showed age-dependent differences we compared the dependency on measurements of the six haemostatic factors II, V, VII, VIII, IX, X by Attard, et al.[1]. Finally, we inspected the data from Attard et al. for differences in the measurements of the six haemostatic factors II, V, VII, VIII, IX, X for different age groups.

ACKNOWLEDGEMENTS
This research was in part conducted and funded by the Australian Research Council Centre of Excellence in Convergent Bio-Nano Science and Technology (project number CE140100036). This study was supported by the Victorian Government's Operational Infrastructure Support Program.

AUTHOR CONTRIBUTIONS
All authors collaboratively conceived the study. IS developed and implemented the modelling approach with important contributions of KL and EC. VI and PM provided the raw experimental data (previously published in aggregated form), and guidance for the interpretation of these data and appropriate representation in the mathematical model. IS prepared all figures and together with SB, drafted the paper. All authors critically revised the manuscript for important intellectual content and approved the final version. VI and EC contributed equally to this study.

COMPETING INTERESTS STATEMENT
The authors declare that they have no conflicts of interest with the contents of this article.

## DATA AVAILABILITY STATEMENT

The datasets analysed during this study are available from the corresponding author upon reasonable request.

## ETHICAL APPROVAL AND INFORMED CONSENT

Ethical approval for the experiments carried out in Attard et al. which this work is based on was obtained as described there. Briefly, the collection of samples from children and adults was approved by the Royal Children's Hospital Ethics in human Research Committee, reference number 2003. The collection of neonatal samples was approved by the Royal Women's Hospital Research Ethics Committee, project 02/08. Informed consent was obtained from the parents of the neonates and children and from the adult participants themselves. All experiments were conducted according to the guidelines of the Declaration of Helsinki.

Tables

**Table 1:** Average, minimum and maximum levels of haemostatic factors for the data from Attard et al. (1). The average adult results are chosen as a standard which is assumed to correspond to the initial concentrations of the model by Hockin et al. (9). For example, the average level of prothrombin 1.33 IU/mL is assumed to equate 1,400 nM. As an example we calculate the mean prothrombin level for neonates: $0.562 / 1.33 \cdot 1{,}400$ nM $\approx 591$ nM. For the individual age group we provide minimum, maximum and mean concentrations for factors II, V, VII, VIII, IX, X so that the range of these concentrations can be assessed.

|  |  | II | | V | | VII | | VIII | | IX | | X | |
|---|---|---|---|---|---|---|---|---|---|---|---|---|---|
|  |  | IU/mL | nM | IU/mL | nM | % | nM | % | nM | % | nM | % | nM |
| Adults | Min | 0.692 | 662 | 0.061 | 5.14 | 29.811 | 3.54 | 43.344 | 0.360 | 47.668 | 65.4 | 42.466 | 52.4 |
|  | **Standard** | **1.33** | **1,400** | **0.237** | **20** | **84** | **10** | **84** | **0.7** | **66** | **90** | **130** | **160** |
|  | Max | 2.08 | 2194 | 0.46 | 39.1 | 145.71 | 17.3 | 155.19 | 1.29 | 92.94 | 128 | 466.76 | 576 |
| day 1 | minimum | 0.070 | 73.1 | 0.053 | 4.47 | 20.635 | 2.45 | 47.125 | 0.392 | 13.121 | 18.0 | 29.884 | 36.9 |
|  | mean | 0.562 | 591 | 0.24 | 20.3 | 38.3 | 4.55 | 102.2 | 0.849 | 25.4 | 34.9 | 47.5 | 58.6 |
|  | maximum | 0.960 | 1010 | 0.475 | 40.1 | 56.595 | 6.71 | 162.156 | 1.35 | 43.158 | 59.2 | 71.286 | 88.0 |
| day 3 | minimum | 0.491 | 517 | 0.188 | 15.9 | 23.012 | 2.73 | 50.710 | 0.421 | 15.282 | 21.0 | 28.360 | 35.0 |
|  | mean | 0.703 | 739 | 0.271 | 22.9 | 42.2 | 5.01 | 84.5 | 0.702 | 33.2 | 45.6 | 56.9 | 70.2 |
|  | maximum | 0.803 | 845 | 0.339 | 28.6 | 52.803 | 6.26 | 127.678 | 1.06 | 59.763 | 82.0 | 85.846 | 106 |
| less than 1 year | minimum | 0.131 | 137 | 0.097 | 8.21 | 40.238 | 4.77 | 39.975 | 0.332 | 32.625 | 44.8 | 28.062 | 34.6 |
|  | mean | 1.277 | 1344 | 0.317 | 26.8 | 68.808 | 8.16 | 83.694 | 0.695 | 44.548 | 61.1 | 69.942 | 86.3 |
|  | maximum | 1.913 | 2013 | 0.449 | 37.9 | 101.139 | 12.0 | 196.999 | 1.64 | 76.509 | 105 | 163.719 | 202 |
| 1-5 years | minimum | 0.721 | 759 | 0.047 | 3.94 | 53.114 | 6.30 | 60.573 | 0.503 | 37.159 | 51.0 | 47.916 | 59.1 |
|  | mean | 1.136 | 1196 | 0.351 | 29.6 | 67.832 | 8.05 | 96.751 | 0.804 | 48.353 | 66.3 | 122.739 | 151 |
|  | maximum | 2.467 | 2596 | 0.891 | 75.2 | 91.835 | 10.9 | 158.822 | 1.32 | 74.047 | 102 | 321.891 | 397 |
| 5-10 years | minimum | 0.652 | 686 | 0.117 | 9.84 | 37.825 | 4.49 | 37.479 | 0.311 | 32.462 | 44.5 | 57.961 | 71.5 |
|  | mean | 1.120 | 1178 | 0.306 | 25.9 | 72.360 | 8.58 | 92.810 | 0.771 | 54.9 | 75.4 | 120.678 | 149 |
|  | maximum | 1.660 | 1747 | 0.583 | 49.2 | 113.635 | 13.5 | 176.977 | 1.47 | 114.873 | 158 | 265.432 | 328 |
| 10-16 years | minimum | 0.869 | 914 | 0.018 | 1.51 | 48.247 | 5.72 | 37.834 | 0.314 | 42.099 | 57.8 | 26.413 | 32.6 |
|  | mean | 1.414 | 1488 | 0.175 | 14.8 | 90.820 | 10.8 | 92.762 | 0.771 | 59.458 | 81.6 | 96.040 | 119 |
|  | maximum | 1.876 | 1974 | 0.354 | 29.9 | 174.035 | 20.6 | 137.973 | 1.15 | 97.266 | 133 | 276.549 | 341 |

Figure Legends:

Figure 1: Age-dependent variability of blood clotting factors (see Attard, et al. [1]). Haemostatic factor abundances are both shown on the original scales in units of IU/mL or percentages % as well as in units of nM obtained by conversion according to Table 1.

Figure 2: Thrombin generation curve simulated using the Hockin-Mann model with parameters corresponding to an individual adult subject. Four indices that were calculated for comparison of TGCs calculated for different subjects are shown; these are lag time (LAG), time to peak (TTP), maximum thrombin concentration (MAX) and area under thrombin curve (AUC). For the TGC plotted here we have LAG=554s, TTP=742s, MAX=570nM and AUC=2191 nM·min.

Figure 3: Age-dependent changes of thrombin generation curve. Whereas lag time (LAG) and time to thrombin peak (TTP) seem not to vary with age, thrombin maximum (MAX) and the area under the thrombin curve (AUC) increase roughly two-fold with age.

Figure 4: The thrombin maximum MAX increases approximately linearly with prothrombin. Similar relationships are seen for the other factors but the variation for the predicted thrombin maxima is much higher than the dependency on prothrombin shown here. The prothrombin level is both shown on the original scale in units of IU/mL from Attard, et al. [1] as well as in units of nM obtained by conversion according to Table 1.

Figure 5: The area under the thrombin curve (AUC) is solely determined by the concentration of prothrombin. The prothrombin level is both shown on the original scale in units of IU/mL from Attard, et al. [1] as well as in units of nM obtained by conversion according to Table 1.

Figure 6: Age-stratified experimental data for the area under the thrombin curve (AUC) is compared with the model. Both data and model simulations are scaled by the mean adult level to demonstrate age-related differences. The data (white) confirms the approximate two-fold relative increase of AUC to the adult level after day 3.

Figures

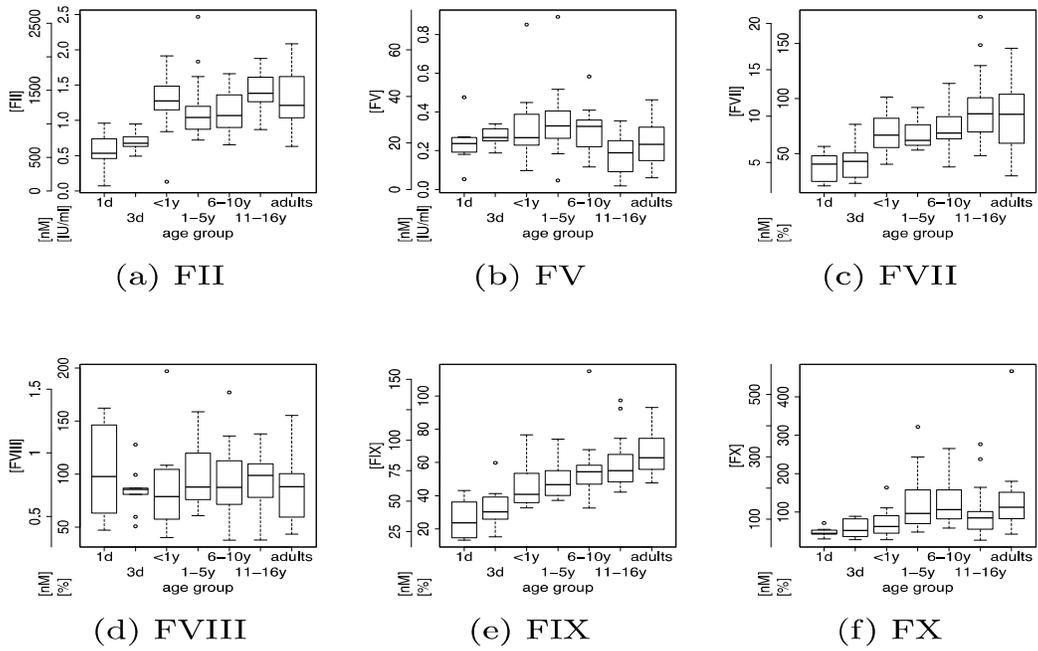

**Figure** 1
13

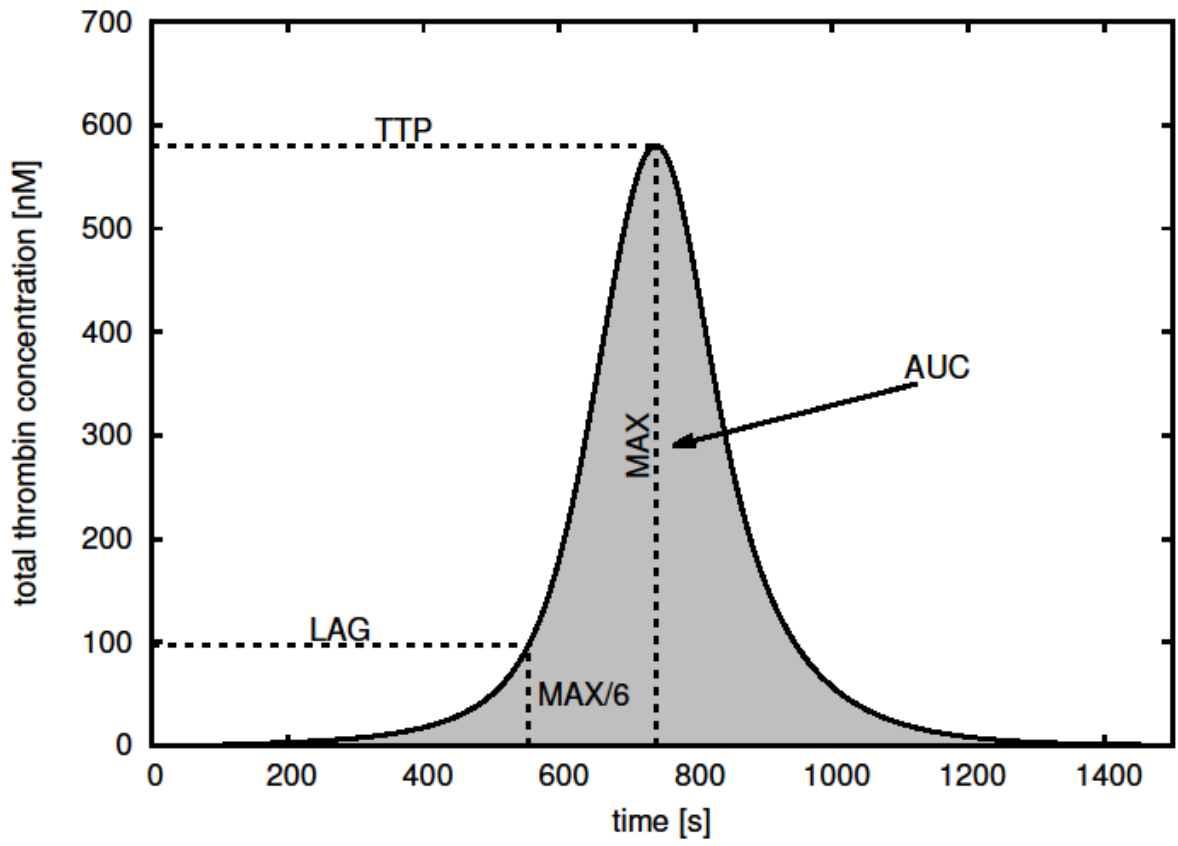

Figure 2

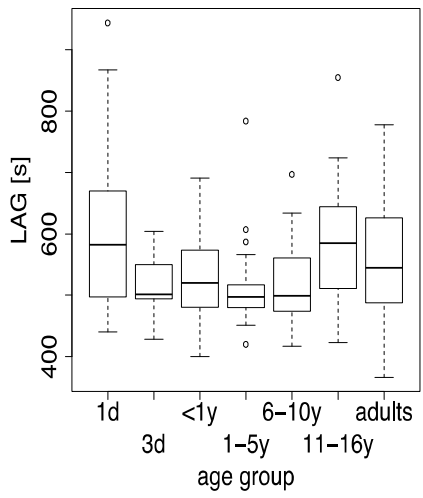
(a) LAG

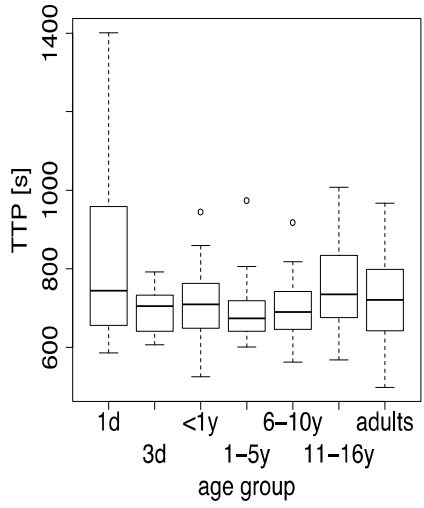
(b) TTP

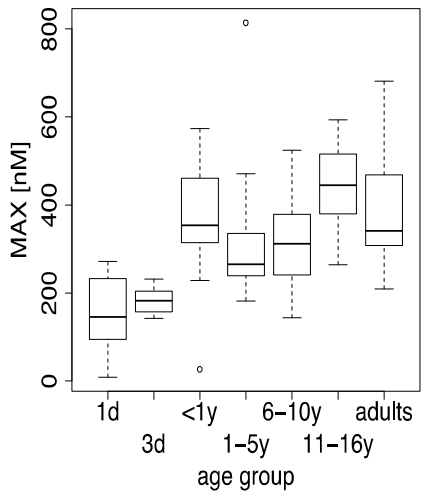
(c) MAX

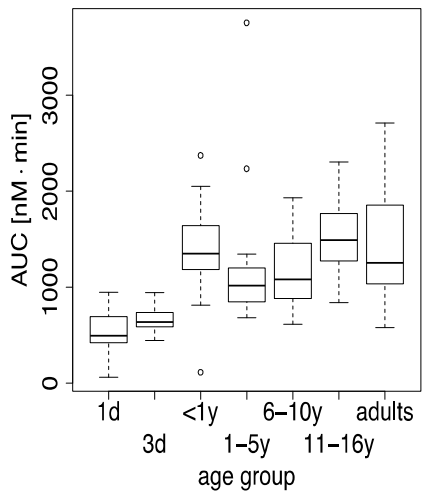
(d) AUC

**Figure 3**

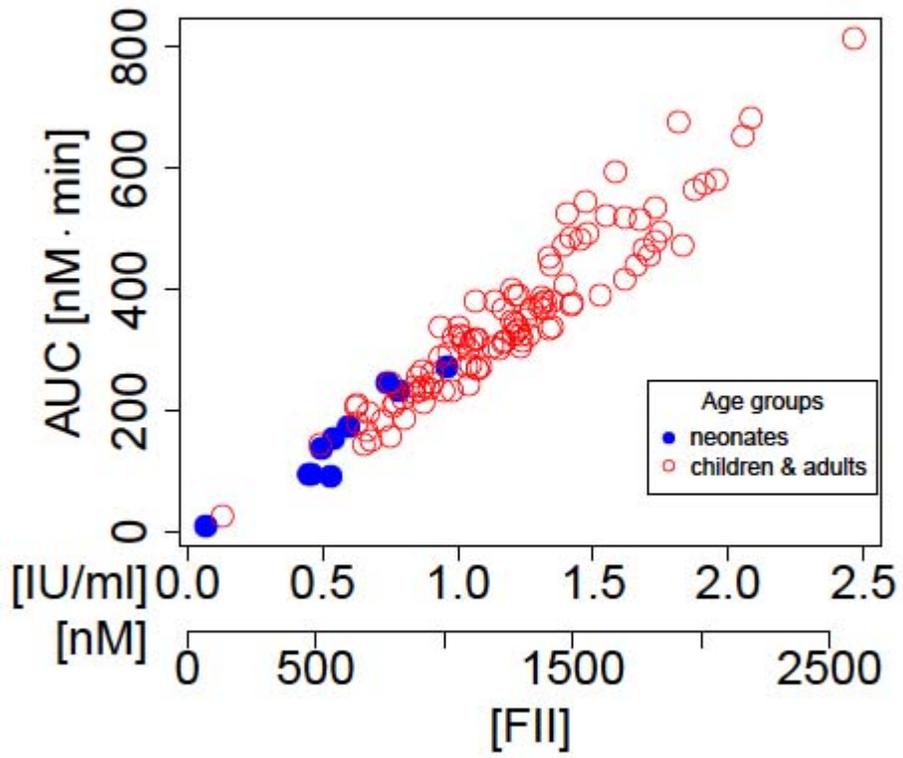

**Figure 4**

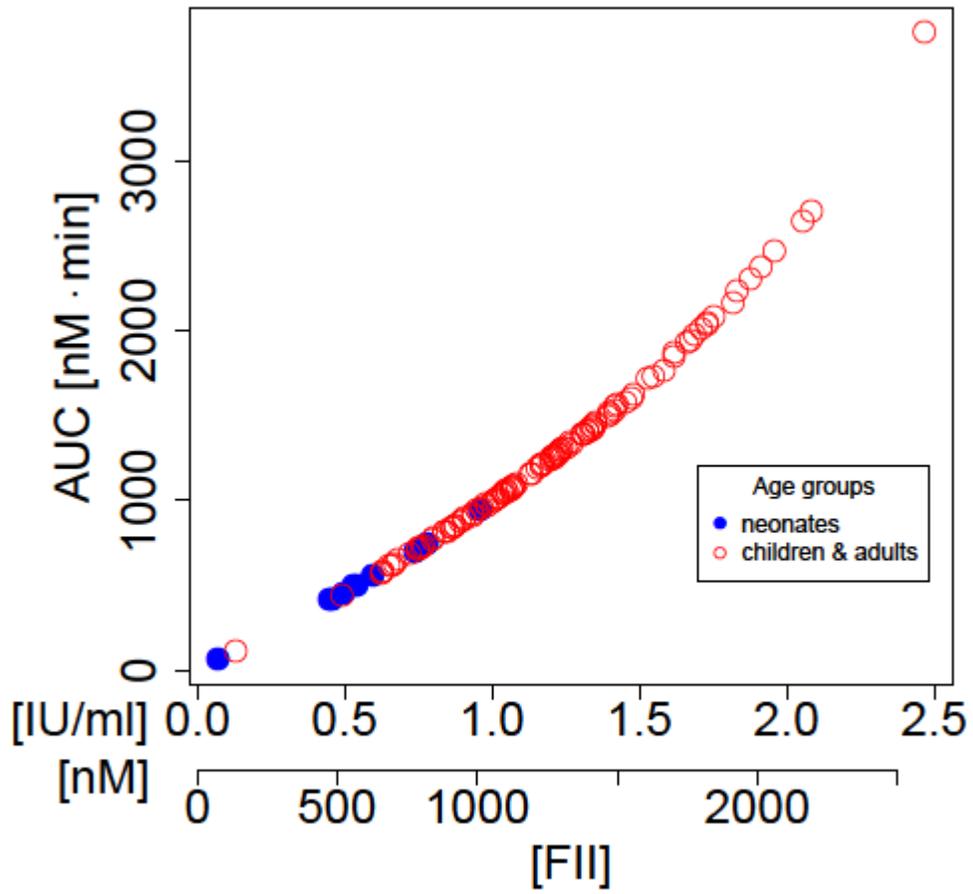

**Figure 5**

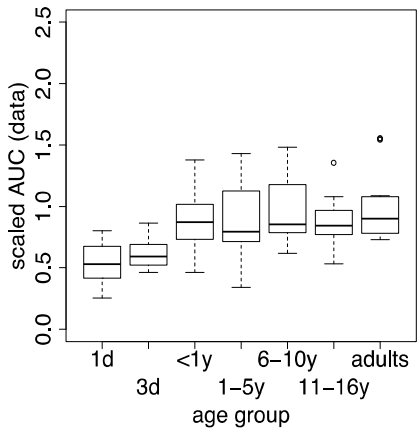 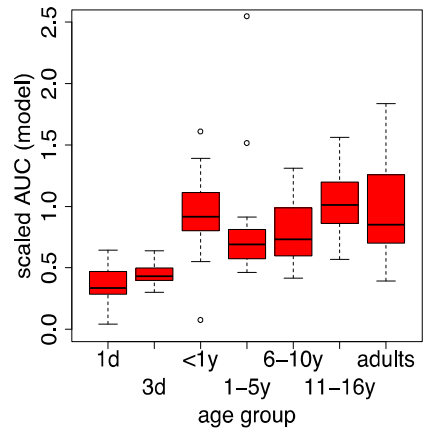

(a) data  (b) model

**Figure 6**